\documentclass[lettersize,journal]{IEEEtran}
     \usepackage{subcaption}
    \usepackage[table]{xcolor}
    \usepackage{amsmath,amssymb,amsfonts}
    \usepackage{mathtools}
    
    \usepackage{algorithmic}
    \usepackage{graphicx}
    \usepackage{textcomp}
    \usepackage{color}
    \usepackage[acronym,shortcuts]{glossaries}
    \usepackage[inline]{enumitem}
    \usepackage{placeins}
    \usepackage[keeplastbox]{flushend}
    \usepackage[utf8]{inputenc}
    \usepackage{tkz-kiviat,pgfplots}
    \pgfplotsset{compat=newest}
    \DeclareUnicodeCharacter{2212}{−}
    \usepgfplotslibrary{groupplots,dateplot}
    \usetikzlibrary{patterns,shapes.arrows}
    \usetikzlibrary{arrows}
    \usetikzlibrary{pgfplots.statistics}
    
    \usepackage{mdframed}
    \usepackage{lipsum}
    \usepackage{booktabs}
    \usepackage[sort&compress, numbers]{natbib}
    \usepackage[binary-units=true]{siunitx}
    \DeclareSIUnit{\belmilliwatt}{Bm}
    \DeclareSIUnit{\dBm}{\deci\belmilliwatt}
    \usepackage[font=footnotesize]{caption}
    \usepackage{float}
    \usepackage{multirow}

    \usepackage{pgfplots}
    \usepackage{setspace}
   
    \usepgfplotslibrary{dateplot} 
    \usepackage{eurosym}
    \usepackage{tabularx}
    \usepackage{multirow}
    
    \usepackage[normalem]{ulem}
    \usepackage{tablefootnote}
    \usepackage{threeparttable}

    \usepackage{scalerel}
    \usetikzlibrary{svg.path}
    \usepackage{hyperref}
    \definecolor{orcidlogocol}{HTML}{A6CE39}
    \tikzset{
      orcidlogo/.pic={
        \fill[orcidlogocol] svg{M256,128c0,70.7-57.3,128-128,128C57.3,256,0,198.7,0,128C0,57.3,57.3,0,128,0C198.7,0,256,57.3,256,128z};
        \fill[white] svg{M86.3,186.2H70.9V79.1h15.4v48.4V186.2z}
                     svg{M108.9,79.1h41.6c39.6,0,57,28.3,57,53.6c0,27.5-21.5,53.6-56.8,53.6h-41.8V79.1z M124.3,172.4h24.5c34.9,0,42.9-26.5,42.9-39.7c0-21.5-13.7-39.7-43.7-39.7h-23.7V172.4z}
                     svg{M88.7,56.8c0,5.5-4.5,10.1-10.1,10.1c-5.6,0-10.1-4.6-10.1-10.1c0-5.6,4.5-10.1,10.1-10.1C84.2,46.7,88.7,51.3,88.7,56.8z};
      }
    }
    \newcommand\orcidicon[1]{\href{https://orcid.org/#1}{\mbox{\scalerel*{
    \begin{tikzpicture}[yscale=-1,transform shape]
    \pic{orcidlogo};
    \end{tikzpicture}
    }{|}}}}
    
    \hypersetup{draft} 

    \definecolor{colorCE0}{HTML}{fc8d59}
    \definecolor{colorDarkCE0}{HTML}{D67147}
    \definecolor{colorCE1}{HTML}{99d594}
    \definecolor{colorDarkCE1}{HTML}{628F5E}
    \definecolor{colorCE2}{HTML}{3288bd}
    \definecolor{colorDarkCE2}{HTML}{286FA2}

    \definecolor{colorSF7}{HTML}{1b9e77}
    \definecolor{colorSF8}{HTML}{d95f02}
    \definecolor{colorSF9}{HTML}{7570b3}
    \definecolor{colorSF10}{HTML}{e7298a}
    \definecolor{colorSF11}{HTML}{66a61e}
    \definecolor{colorSF12}{HTML}{e6ab02}

    \definecolor{colorLoRa}{HTML}{af8dc3}
    \definecolor{colorNBIoT}{HTML}{7fbf7b}
    \definecolor{colorSigfox}{HTML}{fc8d59}

    \makeglossaries
    \newacronym{gprs}{GPRS}{general packet radio services}
\newacronym{egprs}{EGPRS}{enhanced data rates for GSM evolution}
\newacronym{iot}{IoT}{internet of things}
\newacronym{nbiot}{NB-IoT}{narrowband IoT}
\newacronym{lte}{LTE}{long term evolution}
\newacronym[plural=LPWANs,firstplural=low power wide area networks (LPWANs)]{lpwan}{LPWAN}{low power wide area network}
\newacronym{lpwa}{LPWA}{low power wide area}
\newacronym{lorawan}{LoRaWAN}{long range wide area network}
\newacronym{lora}{LoRa}{long range}
\newacronym{nb}{NB}{narrowband}
\newacronym{m2m}{M2M}{machine to machine}
\newacronym{cbm}{CBM}{condition based maintenance}
\newacronym{ai}{AI}{artificial intelligence}
\newacronym{edrx}{eDRX}{extended discontinuous reception mode}
\newacronym{drx}{DRX}{discontinuous reception mode}
\newacronym{psm}{PSM}{power saving mode}
\newacronym{gps}{GPS}{global positioning system}
\newacronym{ptw}{PTW}{paging time window}
\newacronym{tau}{TAU}{tracking area update}
\newacronym{mtc}{MTC}{machine-type communication}
\newacronym{ue}{UE}{user equipment}
\newacronym{cdrx}{CDRX}{connected mode drx}
\newacronym{3gpp}{3GPP}{3rd generation partnership project}
\newacronym{gsm}{GSM}{global system for mobile communications}
\newacronym{rf}{RF}{radio frequency}
\newacronym{prbs}{PRBs}{physical resource blocks}
\newacronym{mcl}{MCL}{maximum coupling loss}
\newacronym{ce}{CE}{coverage enhancement}
\newacronym{rsrp}{RSRP}{reference signals received power}
\newacronym{toa}{ToA}{time on air}
\newacronym{rrc}{RRC}{radio resource connection}
\newacronym{rtc}{RTC}{real time clock}
\newacronym{css}{CSS}{chirp spread spectrum}
\newacronym{ism}{ISM}{industrial, scientific and medical}
\newacronym{mmtc}{MMTC}{massive machine type communication}
\newacronym{sf}{SF}{spreading factor}
\newacronym{cr}{CR}{coding rate}
\newacronym{bw}{BW}{bandwidth}
\newacronym{adr}{ADR}{adaptive data rate}
\newacronym{abp}{ABP}{authentication by personalisation}
\newacronym{otaa}{OTAA}{over the air authentication}
\newacronym{ofdm}{OFDM}{orthogonal frequency division multiplexing}
\newacronym{scfdma}{SCFDMA}{single-carrier frequency division multiple access}
\newacronym{udp}{UDP}{user datagram protocol}
\newacronym{qos}{QoS}{quality of service}
\newacronym{ra}{RA}{random access}
\newacronym{rar}{RAR}{random access response}
\newacronym{snr}{SNR}{signal-to-noise ratio}
\newacronym{ota}{OTA}{over the air}
\newacronym{ttm}{TTM}{time to market}
\newacronym{cots}{COTS}{commercial off-the-shelf}
\newacronym{mac}{MAC}{medium access control}
\newacronym{multi-rat}{multi-RAT}{multiple radio access technology}
\newacronym{rat}{RAT}{radio access technology}
\newacronym{pcb}{PCB}{printed circuit board}
\newacronym{wsn}{WSN}{wireless sensor network}
\newacronym{dbpsk}{DBPSK}{differential binary phase-shift keying}
\newacronym{unb}{UNB}{ultra-narrow band}
\newacronym{soc}{SoC}{state of charge}
\newacronym{rssi}{RSSI}{received signal strength indicator}
\newacronym{sinr}{SINR}{signal-to-interference-plus-noise ratio}
\newacronym{rsrq}{RSRQ}{reference signal received quality}
\newacronym{gnss}{GNSS}{global navigation satelite system}
\newacronym{pdr}{PDR}{packet delivery ratio}
\newacronym{tcp}{TCP}{transmission control protocol}
    \setcitestyle{square}
    \def\BibTeX{{\rm B\kern-.05em{\sc i\kern-.025em b}\kern-.08em
        T\kern-.1667em\lower.7ex\hbox{E}\kern-.125emX}}

    \newcommand{\lievenok}[1]{}
    \newcommand{\liesbetok}[1]{}
    \newcommand{\gillesok}[1]{}
    \newcommand{\geofok}[1]{}
    \newcommand{\guusok}[1]{}

    \newcommand{\old}[1]{{}}
    \newcommand{\ie}{{i.e.,\ }}
    \newcommand{\eg}{{e.g.,\ }}
    
\begin{document}

\title{Multi-RAT IoT -- What's to Gain?\\ An Energy-Monitoring Platform}



\author{
    \IEEEauthorblockN{Guus Leenders \orcidicon{0000-0001-9633-2584}, Gilles Callebaut \orcidicon{0000-0003-2413-986X},  Liesbet Van der Perre \orcidicon{0000-0002-9158-9628}, Lieven De Strycker \orcidicon{0000-0001-8172-9650}\\}
    \IEEEauthorblockA{
        \textit{KU Leuven, ESAT-DRAMCO, Ghent Technology Campus}\\
        Ghent, Belgium\\
        name.surname@kuleuven.be
    }
}

\maketitle

\begin{abstract}
Multiple \acrshortpl{lpwan} have been rolled out to support the variety of \acrshort{iot} applications that are crucial to the ongoing digital transformation. These networks vary largely in terms of quality-of-service, throughput and energy-efficiency. To cover all \acrshort{lpwan} use-cases most optimally, multiple networks can be combined into a \gls{multi-rat} solution. In particular environmental monitoring in both smart city and remote landscapes. We present and share such a \acrshort{multi-rat} platform. To derive an accurate profile of the \acrshort{multi-rat} opportunities in various scenarios, in the-field network parameter are monitored. The platform collects per-packet energy-consumption, \gls{pdr} and other parameters of LoRaWAN, NB-IoT and Sigfox. 
Our preliminary measurements demonstrate the validity of using a \acrshort{multi-rat} solution.  For example, we illustrate the potential energy savings when adopting \acrshort{multi-rat} in various scenarios.

\end{abstract}

\section{Introduction}
\glsunset{lorawan}

\Gls{iot} networks are being deployed for a great diversity of use cases. Many of them feature environmental monitoring applications. Monitoring air pollution in smart city landscapes~\cite{candia2018solutions} is a prime example. Another application is the monitoring of tree growth and soil moisture levels in remote rural areas~\cite{matasov2020iot}.  Herein, a vast \gls{wsn} is deployed: connecting \gls{iot} sensor nodes to the Internet. These devices require long-range and low-power connectivity, limiting the wireless interface to \gls{lpwan} technologies.

\begin{figure}[t]
    \centering
    \includegraphics[width=.38\textwidth]{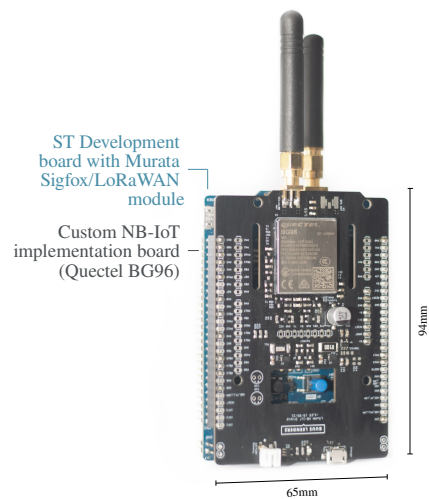}
    \caption{Picture of the presented, custom and open-source \gls{multi-rat} platform~\cite{opensourcesoftware} to perform per-packet energy measurements of \gls{iot} communication in \gls{nbiot}, \gls{lorawan}, and Sigfox networks. 
    }\label{fig:picture}
\end{figure}

\Gls{iot} devices are typically battery-powered, so extra design efforts are needed to optimize the energy usage of \gls{iot} nodes~\cite{callebaut2021art}. In this paper, we consider \gls{lorawan}, Sigfox and \gls{nbiot}. These technologies feature long-range connectivity whilst remaining relatively low power~\cite{mekki2019comparative}. They vary, however, largely in terms of data bandwidth, \gls{qos} and energy-efficiency~\cite{li2017smart, mekki2019comparative}. For example, while \gls{nbiot} provides more data bandwidth and \gls{qos}, its energy efficiency is considerably worse than Sigfox and LoRaWAN, especially when transmitting smaller payloads~\cite{leenders2021multi}. Due to these inherently different characteristics between \gls{lpwan} solutions, it is challenging to optimize any single technology for energy-efficiency for a more complex and varying use case. It is clear that these differences can be exploited by combining multiple wireless connectivity solutions in one \gls{iot} node. 







By combining multiple wireless technologies into one \gls{multi-rat} \gls{iot} node~\cite{leenders2021multi}, several crucial benefits can be obtained: 
\begin{enumerate*}[label=(\arabic*)]
    \item energy-efficient operation for variable payload sizes,
    \item timely delivery for latency-critical messages,
    \item improved service area, and
    \item improved \gls{qos}.
\end{enumerate*}
The energy consumption of the IoT node is determined by various parameters, both network and environment related. 
Literature focuses mainly on theoretical energy models or models obtained in lab conditions~\cite{michelinakis2020dissecting}, omitting the non-negligible impact caused by vendor and operator-specific configurations or various coverage conditions. One publication by \citeauthor{michelinakis2020dissecting}, published in-the-field observations of large differences in energy consumption between various network providers of \gls{nbiot}.
This highlights the need for platforms to obtain an accurate comparison and energy model  of \gls{lpwan} \gls{iot} technologies, where both network and environment parameters are closely monitored.

\begin{figure*}[t]
    \centering
    \includegraphics[width=.7\textwidth]{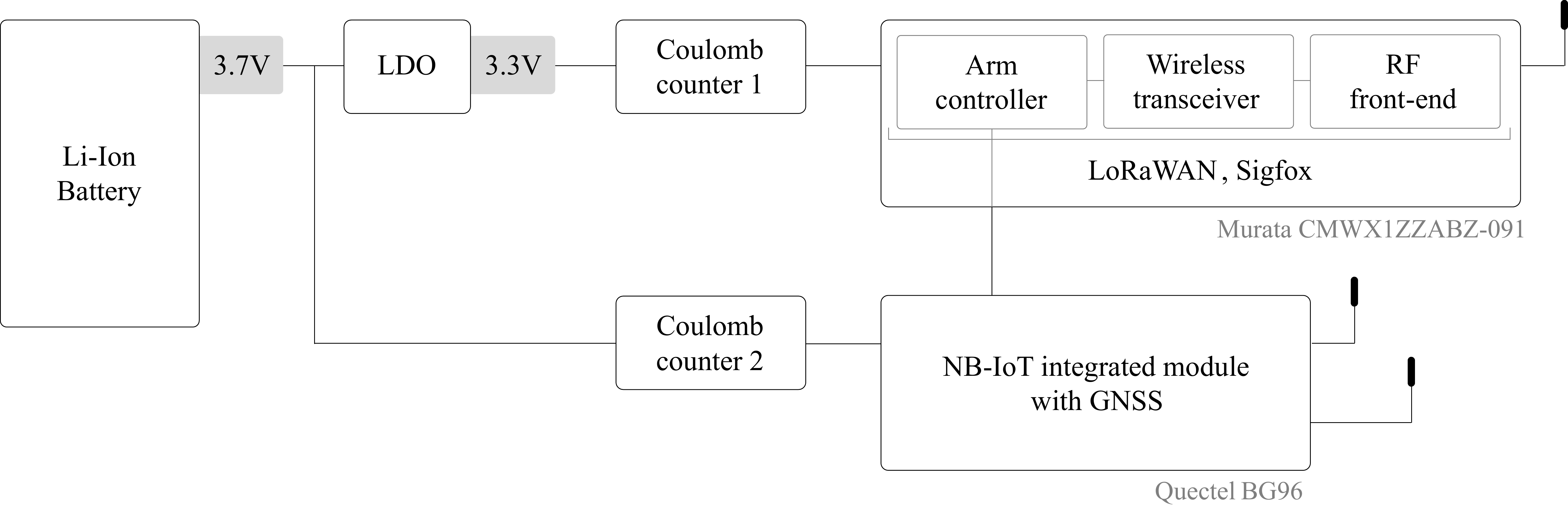}
    \caption{Schematic overview of the proposed system.}
    \label{fig:systemoverview}
\end{figure*}

{\textbf{\textit{Contributions}--\ }} In this work, we present a platform for diverse, in-the-field parameter monitoring for \gls{lpwan} \gls{iot} networks. It includes for instance the power consumption on a per-packet basis. The platform consists of a custom-made hardware measuring module, as well as a cloud-based interface with an easy to interpret dashboard. Through this platform, we demonstrate  the shortcomings of using a single technology and study the potential of \gls{multi-rat} in typical \gls{iot} settings and use-cases. Moreover, the cloud platform is publicly accessible\footnote{Openly accessible via \url{dramco.be/multi-rat}} and both the software and hardware are open-source~\cite{opensourcesoftware}.

This paper is organized as follows. Firstly, we elaborate on the available wireless \gls{iot} technologies, focusing on various advantages of a \gls{multi-rat} solution.  Secondly, we present our custom-designed \gls{multi-rat} measurement platform. Thirdly, the viability of such measurement platform is demonstrated. Finally, we summarize our main findings and suggest some further platform improvements \eg remotely updating the monitoring algorithm to specific use cases.

\section{Wireless  Technologies and Multi-RAT opportunities}

The considered \gls{lpwan} technologies (\ie \gls{lorawan}, Sigfox and \gls{nbiot}) are tailored to provide wide area coverage to energy-constrained devices. 
However, they differ largely in the design of both the physical layer and medium access control layer~\cite{callebaut2021art}. These differences result in a diverse set of corresponding applications and use cases.  

\subsection{License-exempt technologies.} 
To comply with limitations imposed on the \gls{ism} band, duty cycle limits are put in place~\cite{sornin2015lorawan, aernouts2018sigfox}. This intereference mitigation measure results in a non-negligible minimum delay between packets and limited peak throughput. 

\textit{\Gls{lorawan}} makes use of the, on \gls{css} based, \gls{lora} modulation technique. This technology operates in the  
\gls{ism} band and does not employ any multiple-access technique, \ie ALOHA is used. \gls{lorawan} coverage can be freely extended with private gateways. 

\textit{Sigfox} uses \gls{unb} modulation with \gls{dbpsk} in the same license free \gls{ism} band as \gls{lorawan}~\cite{gomez2019sigfox}. Sigfox is a proprietary technology, and its network is deployed privately. To maximize \gls{pdr}, 
Sigfox will send each data packet three times, each at another random carrier frequency. This diversity technique 
increases the probability of receiving at least one packet successfully. This, however, results in a larger \gls{toa}, thereby further reducing the throughput due to the duty cycle regulations~\cite{aernouts2018sigfox}.

\subsection{Cellular Technology.} 
\textit{\Gls{nbiot}}, as a derivative of LTE, typically operates in the licensed band spectrum.  
Radio resources are allocated by the network to \gls{iot} nodes, based on  time-frequency slots. By simplifying some main LTE principles, a licensed \gls{lpwan} technology is created~\cite{ratasuk2016overview}. 
Special attention was paid to, among other things, lowering hardware complexity, coverage enhancement, and lowering energy consumption. By introducing \gls{edrx} and longer \gls{psm} delays, active radio time is reduced, thereby lowering the overall energy consumption of the \gls{iot} node~\cite{liberg2017cellular}. 

\subsection{Multi-RAT opportunities.}
It is clear that, no matter how widely applicable any of the aforementioned technologies may be, there is no one-fits-all technology. Equipping a node with multiple
wireless technologies, will facilitate switching between message transmissions 
to obtain a higher energy-efficiency or to address new application requirements or context-switches.

When deploying energy-constricted \gls{iot} nodes requiring large payloads, for example, \gls{nbiot} is the prime candidate. Yet, the energy cost of sending periodical `alive' messages is rather high, in comparison to \gls{lorawan} or Sigfox. Therefore, a typical smart city setup, with both \gls{nbiot} and \gls{lorawan}, can reduce the energy consumption by a factor of~4~\cite{leenders2021multi} or more.
Furthermore, \gls{nbiot} power consumption is highly dependent on the operating conditions, such as network coverage and configuration~\cite{michelinakis2020dissecting}. Including alternative wireless radios can increase the energy-efficiency when detecting high energy costs in one network.  
These trade-offs and energy gains can be measured by the presented \gls{multi-rat} platform.

\section{Multi-RAT Platform}

In order to quantitatively assess the potential of a \gls{multi-rat} solution, and investigate possible practical energy efficiency optimization approaches, we created a prototype that is able to map the characterizations of \gls{iot} wireless networks. The realized prototype includes \gls{lorawan}, Sigfox and \gls{nbiot} and is able to measure energy consumption on-device on a per-packet-basis. These measurements are wirelessly transmitted and evaluated in the back-end. This prototype enables \gls{iot} developers to adapt application requirements and restrictions based on measured, on-site energy consumption data.
The prototype is depicted in Fig.~\ref{fig:picture}.  

\subsection{On-Board Wireless Technologies}

A schematic overview of the prototype testing platform is depicted in~Fig.~\ref{fig:systemoverview}. The core of the device features the CMWX1ZZABZ-091 module by Murata. The module is powered by an ultra-low-power Arm Coretx-M0+  microcontroller (STM32L072CZ) and wireless transceiver (SX1276)~\cite{stdatasheet} (\gls{lorawan} and Sigfox). The interface to program the microcontroller is user accessible, contradictory to other wireless modules. Thus, eliminating the need for an extra microcontroller~\cite{callebaut2021art}. The embedded wireless transceiver can be used to connect to either the \gls{lorawan} or Sigfox network. In this work, the controller firmware was customized in such a way, that dynamic switching between both networks is possible.  The CMWX1ZZABZ-091 module hardware is extended with \gls{nbiot} hardware: a custom designed extension board featuring a Quectel~BG96.

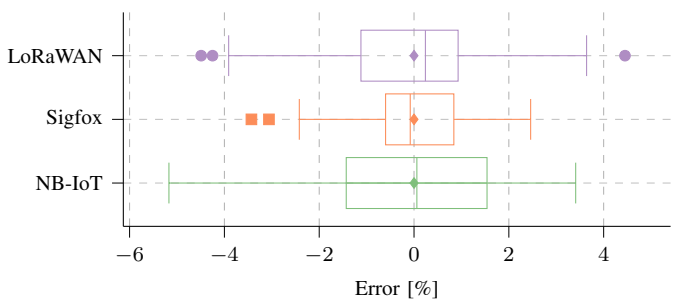
\begin{figure}[b]
    \centering
    \begin{tikzpicture}
  \begin{axis}
    [
    axis line style={black},
    axis lines* = {left},
    legend cell align={left},
    legend style={ at={(0.5,-0.3)},anchor=north, fill opacity=0.8, draw opacity=1, text opacity=1, draw=white!80.00000!black, /tikz/column 2/.style={column sep=5pt}, /tikz/column 3/.style={column sep=5pt},/tikz/column 4/.style={column sep=5pt}},
    legend style={nodes={scale=0.7, transform shape}},
    legend columns=4, 
    tick align=outside,
    tick pos=left,
    width = \linewidth,
    height=0.5\linewidth,
    x grid style={white!69.01960784313725!black},
    grid,
    grid style={help lines,color=gray!50, dashed},
    xtick style={color=black},
    xlabel={Error [\%]},
    ytick style={color=black},
    label style={font=\footnotesize},
    tick label style={font=\footnotesize},
    ytick={1,2,3},
    yticklabels={NB-IoT, Sigfox, LoRaWAN},
    ]
    
    \addplot+ [colorNBIoT,mark options={fill=colorNBIoT},
    boxplot prepared={
      median=0.06,
      upper quartile=-1.43,
      lower quartile=1.54,
      upper whisker=3.41,
      lower whisker=-5.17,
      average=0
    },
    ] coordinates {};
    
    \addplot+ [colorSigfox,mark options={fill=colorSigfox},
    boxplot prepared={
      median=-0.08,
      upper quartile=0.84,
      lower quartile=-0.6,
      upper whisker=2.46,
      lower whisker=-2.42,
      average=0
    },
    ] coordinates {(2,-3.06)(2,-3.43)};
    
    \addplot+ [colorLoRa,mark options={fill=colorLoRa},
    boxplot prepared={
      median=0.24,
      upper quartile=0.93,
      lower quartile=-1.12,
      upper whisker=3.64,
      lower whisker=-3.91,
      average=0
    },
    ] coordinates {(3,-4.49)(3,-4.25)(3,4.45)};

  \end{axis}
\end{tikzpicture}
    \vspace{-0.5cm}
    \caption{Recorded error margin on energy measurements using the proposed system, after calibration. The errors correspond to the energy difference measured by the platform and the Otti~Arc power analyzer. 
    }
    \label{fig:boxplot}
\end{figure}

\subsection{Power/energy measuring}

Power distribution of the platform is illustrated in Fig.~\ref{fig:systemoverview}. 
To accurately measure energy consumption on a per-packed-basis, high accuracy coulomb counting is used. To do so, two coulomb counting modules are used, measuring the power usage on each power rail. The coulomb counting algorithm is performed by the LTC2941 battery gas gauge. 
The accuracy of the calculated energy consumption is compared to a dedicated energy meter instrument, i.e., an Otii~Arc\footnote{More information on Otii~Arc: \url{qoitech.com/otii/}}.
To eliminate an energy offset, the platform is calibrated based on the energy usage reported by both the coulomb counter and Otti~Arc for \num{120}~packets for each \gls{iot} technology. The utilized coulomb counter employs a sense resistor, measuring the voltage over the resistor to determine the current. The difference between the advertised resistance and the real value is determined based on the above-mentioned procedure. By scaling all measurements of the coulomb counter, this error is removed. 

The residual error after calibration is shown in Fig.~\ref{fig:boxplot}. Good overall accuracy is achieved with this low-complexity design. However, the fast transient currents of the \gls{iot} communication result in larger error margins.  These need to be taken into account when determining the energy consumption of each technology. 

\subsection{Parameter Monitoring}
The energy consumption is determined by the configuration parameters of each technology and the context (\eg network coverage). In order to map the measured energy expenditure to these parameters, a detailed list of all actual applicable parameters is retrieved and transmitted to the back-end for further analysis. 

Common parameters for \gls{lorawan}, Sigfox and \gls{nbiot} are:
\begin{itemize}
    \item number of received messages at the gateway/back-end,
    \item number of transmitted messages,
    \item payload size,
    \item time of reception,
    \item location of the receiving gateway(s)/base station(s),
    \item \gls{rssi},
    \item \gls{snr},
    \item transmit power,
    \item \gls{gnss} position,
    \item motion speed.
\end{itemize}

Aside from more general information, a large number of \gls{iot} technology specific data is obtained from the on-board parameters and the available information at the back-end.

For a \gls{lorawan} transmission, the current \gls{sf} is recorded. The evolution of both the transmit power and the \gls{sf} indicate the available \gls{lorawan} coverage when \gls{adr} is enabled. 
For Sigfox transmissions, the estimated region where the device is located, reported in the back-end, is collected. 
For all \gls{nbiot} transmissions, the used \gls{ce} level is stored, summarizing the network coverage. Other transmission specific data is also stored: \gls{rsrp}, \gls{sinr}, and \gls{rsrq}.  \gls{edrx} and \gls{psm} settings are recorded, to estimate the current power profile. 



\subsection{Monitoring Algorithm}
\begin{figure}
    \centering
    \includegraphics[width=.35\textwidth]{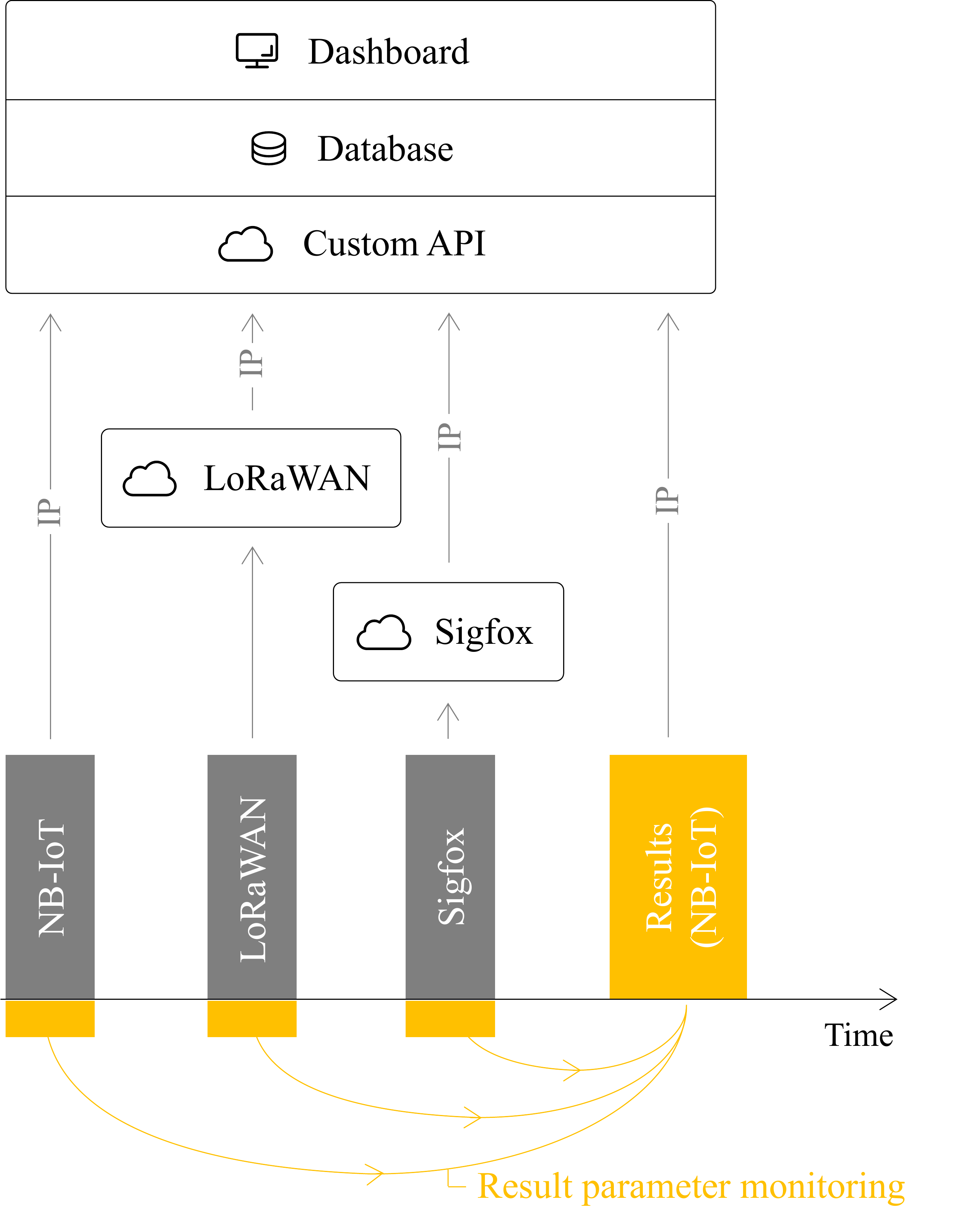}
    \caption{Monitoring algorithm overview.}
    \label{fig:testoverview}
\end{figure}
The aforementioned on-board parameters and energy consumption per technology are acquired according to the approach depicted in Fig.~\ref{fig:testoverview}.
The measurement cycle consists of two parts, where first we transmit one packet per \gls{iot} technology, followed by a message containing the captured results.
In the first phase, both the payload size and transmit power are varied per measurement cycle. The used transmit power depends on \gls{adr} in \gls{lorawan} and the network in \gls{nbiot}. For Sigfox, the transmit power is fixed to \SI{14}{\dBm}. In every measurement cycle, a random payload size is selected per technology. The maximum payload size is different for each technology, \ie 12, 256 and \SI{1547}{\byte} for Sigfox, LoRaWAN and \gls{nbiot}, respectively.

Power consumption is monitored from the very start until the very end of the wireless transmission. In \gls{nbiot} especially, this includes \gls{edrx} or delayed \gls{rrc} releases. Each of these packets get sent to our custom cloud platform, either directly (\gls{nbiot}) or through network operated cloud platforms (\gls{lorawan} and Sigfox). Power consumption and various other previously discussed transmission parameters are recorded and sent to the custom cloud platform using an extra \gls{nbiot} packet. This transmission is not included in the energy consumption metrics and is only used to communicate the stored measurements and configurations. 


\subsection{Dashboard}
To ease the evaluation of the experimental data, a web interface for accessing real-time results was developed. Any relation between the discussed parameters (including consumed energy) can be analyzed by dynamically generating various graphs (\eg scatter plots or maps). By applying multiple filters, one can easily focus on a particular use case. 
Hereby allowing \gls{iot} developers and researches to draw fast and easy conclusions regarding the influence of certain parameters on the energy consumption.

\section{Evaluation}

\begin{table*}[t]
\footnotesize
\centering
\caption{Comparison between \gls{nbiot}, \gls{lorawan} and Sigfox as experimental validation of the presented \gls{multi-rat} platform. \acrfull{pdr} and the average energy per byte $E_b$ serve as comparison metrics.  }
\label{tab:results}
\begin{tabular}{@{}llrrrrrrrrr@{}}

\toprule
                                                                \multirow{3}{1.2cm}{Payload Size (\si{\byte})}    &                                     & 
                                                                    \multicolumn{6}{c}{Static}                                    & \multicolumn{3}{c}{Mobile}       \\ \cmidrule(lr){3-8} 
                                                                    &                                     & \multicolumn{3}{c}{Indoor}      & \multicolumn{3}{c}{Outdoor} & \multicolumn{3}{c}{Outdoor}      \\ \cmidrule(lr){3-5} 
                                          \cmidrule(lr){6-8}
                                                                \cmidrule(lr){9-11}
 &                                     & NB-IoT & LoRaWAN       & Sigfox & NB-IoT  & LoRaWAN  & Sigfox & NB-IoT & LoRaWAN        & Sigfox \\ \midrule
1-12                                                                & PDR (\%)                            & 93.10   & 61.58         & 89.86  & 94.54        & 52.89         & 73.49       & 88.89  & 62.09          & 42.98  \\
                                                                    & $E_b$ (\si{\micro\watt\hour\per\byte}) & 60.52  & 
                                                                    8.03          & 45.58  & 44.36        &  11.65        & 47.03       & 74.80   & 10.2           & 50.79  \\
                                                                    \\
12-51                                                               & PDR (\%)                            & 98.53  & 71.90          & -      &  92.85       & 53.95         &  -      & 81.98  & 58.46          & -      \\
                                                                    & $E_b$ (\si{\micro\watt\hour\per\byte}) & 12.61  & 3.69          & -      & 18.65         & 6.56         & -       & 32.85  & 0.53           & -      \\
                                                                    \\
51-255                                                              & PDR (\%)                            & 97.17  & 72.00  & -      & 92.89        & /         &  -      & 84.78  & /   & -      \\
                                                                    & $E_b$ (\si{\micro\watt\hour\per\byte}) & 5.98   & 0.33 & -      & 3.95        &  /        & -       & 10.12  & / & -      \\
                                                                    \\
255-1547                                                                & PDR (\%)                            & 99.08  & -             & -      & 90.63        & -        & -      & 82.86  & -              & -      \\
                                                                    & $E_b$ (\si{\micro\watt\hour\per\byte}) & 1.03   & -             & -      & 0.81        & -        & -      & 0.89   & -              & -      \\
\bottomrule
\end{tabular}
\end{table*}

The proposed platform is experimentally validated by mapping the characteristics of several \gls{iot} networks in various circumstances (location, moving speed, indoor/outdoor and urban/rural). These circumstances correspond to a large number of use cases: performing sensor ratings in smart cities at a fixed location, tracking the movement of  (rental) bikes in cities or monitoring environmental tracking of sensitive assets during transport. 

In this validation, we focus on two parameters: \gls{pdr} and energy consumption per (payload) byte.  The \gls{pdr} and energy consumption are measured for every individual \gls{iot} technology. 
In the experiments, any confirmation of reception is disabled on all \gls{iot} technologies. Downlink parameters are enabled, though, only to enable energy optimizing strategies such as \gls{adr} in \gls{lorawan}.

These experimental results were collected at various locations across the Belgian region of Flanders, covering an area of  \SI{3000}{\kilo\meter\squared}. 
Mobile measurements are collected by means of commuting between two cities using bike and train. In total, \num{4597} data points are gathered.
The networks used are the Proximus \gls{nbiot} and \gls{lorawan} network.  CityMesh is the official Sigfox Operator in this region. All of these operators claim full coverage across the region. 

\begin{figure}[H]
    \centering
    \begin{tikzpicture}
  \begin{axis}[
    axis line style={black},
    axis lines* = {left},
    legend cell align={left},
    legend style={ at={(0.5,-0.3)},anchor=north, fill opacity=0.8, draw opacity=1, text opacity=1, draw=white!80.00000!black, /tikz/column 2/.style={column sep=5pt}, /tikz/column 3/.style={column sep=5pt},/tikz/column 4/.style={column sep=5pt}},
    legend style={draw=none, nodes={scale=0.7, transform shape}},
    legend columns=4, 
    tick align=outside,
    tick pos=left,
    width = \linewidth,
    height=0.49\linewidth,
    x grid style={white!69.01960784313725!black},
    grid,
    grid style={help lines,color=gray!50, dashed},
    xtick style={color=black},
    ylabel={PDR [\%]},
    ytick style={color=black},
    label style={font=\footnotesize},
    tick label style={font=\footnotesize},
    xtick={1,2,3,4},
    xticklabels={Static, $<$10 km/h, 10-30\,km/h, $>$30\,km/h},
    ]
    
    \addplot[colorNBIoT, thick] coordinates
      {(1,88) 
      (2,83) 
      (3,86) 
      (4,79)};
      \addlegendentry{NB-IoT}
      
    \addplot[colorLoRa, thick] coordinates
      {(1,51) 
      (2,51) 
      (3,56) 
      (4,43)};
      \addlegendentry{LoRaWAN}
    
    \addplot[colorSigfox, thick] coordinates
      {(1,78) 
      (2,53) 
      (3,34) 
      (4,17) };
      \addlegendentry{Sigfox}
  \end{axis}
\end{tikzpicture}
    \caption{Comparison of \acrfull{pdr} of \gls{nbiot}, \gls{lorawan} and Sigfox in function of speed of the \gls{iot} node for packet payloads of \SI{1}{\byte}-\SI{12}{\byte}. 
    }
    \label{fig:mobility}
\end{figure}
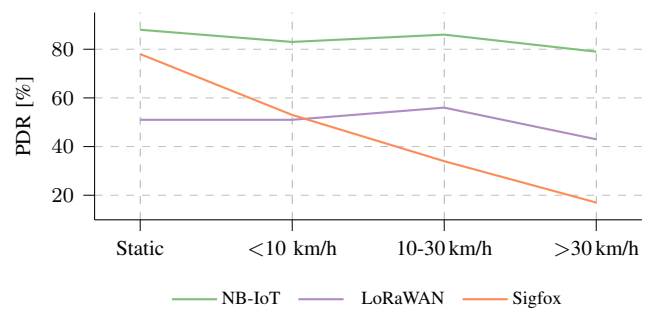

The results of the conducted experimental campaign are summarized in Table~\ref{tab:results} and provide insight in both the \gls{pdr} and energy consumption in various scenarios and various payload sizes. Based on this data, we derive three \gls{multi-rat} conclusions or opportunities. 
\begin{enumerate}
    \item \textit{Payload dependent \gls{multi-rat} switching for efficient energy  consumption.} 
    \gls{iot} use cases with varying payload sizes will benefit greatly in terms of energy consumption from implementing a \gls{multi-rat} scheme. By only employing \gls{nbiot} for sending larger messages (\SI{51}{\byte} - \SI{1547}{\byte}), and using \gls{lorawan} for smaller messages (\SI{1}{\byte} - \SI{51}{\byte}), energy is saved. As seen in Table~\ref{tab:results}, energy consumption per byte ($E_b$) for smaller messages (\SI{1}{\byte} - \SI{51}{\byte}), improves slightly when comparing Sigfox to \gls{nbiot} but at least quadruples when comparing \gls{lorawan} to \gls{nbiot}. With the inclusion of \gls{nbiot}, large packets can still be sent without the need for splitting payloads across different \gls{lorawan} or Sigfox packets (and thus otherwise increasing latency).
    
    \item \textit{Mobility management}
    \gls{iot} communication on mobile nodes can suffer from low \gls{pdr} when moving at high speed. 
    By comparing static to mobile measurements, it is clear that the \gls{pdr} of Sigfox dramatically diminishes for mobile nodes. This can also be seen when plotting the \gls{pdr} versus speed in the measurement platform dashboard (Fig.~\ref{fig:mobility}). For Sigfox, \gls{pdr} drops with increasing speed, while the \gls{pdr} of both \gls{nbiot} and \gls{lorawan} largely remains constant. 
    These results are consistent with measurements gathered by \citeauthor{wang2020performance}~\cite{wang2020performance}, in which packets are transmitted from \gls{iot} nodes moving at high speed.

    \item \textit{Energy efficient \gls{multi-rat} switching for mission-critical \gls{iot}.}
    Guaranteeing delivery in \gls{iot} requires a downlink channel for confirmation packets to be sent. Traditionally, one would opt for a high \gls{pdr} \gls{iot} technology such as \gls{nbiot}. However, the energy cost of sending a \gls{nbiot} packet is relatively high when compared to \gls{lorawan} and Sigfox. First, attempting to send critical data via \gls{lorawan} or Sigfox (with a confirmation downlink channel) can improve energy consumption drastically. 
    
\end{enumerate}

\FloatBarrier
\section{Conclusion}
In this work, we presented a \gls{multi-rat} platform for in the field analysis of multiple \gls{iot} networks, with a particular focus on energy consumption in real life scenarios such as remote monitoring of environmental parameters. A diverse set of parameters is monitored (such as \gls{rssi} or location) to  be able to match energy consumption with various environmental circumstances.  All data relations can be studied via an easy-to-use web interface. 
When developing \gls{iot} solutions for highly specific use cases, the proposed platform can be used to determine  \gls{multi-rat} insights and optimization before large-scale adoption. 
As experimental evaluation of the proposed platform, we studied the \gls{pdr} and energy consumption per byte $E_b$ and posed three \gls{multi-rat} conclusions for common use cases. 
This platform opens many opportunities for further research. For example, an analysis on the impact of confirmed communication of \gls{iot} nodes is required, most notably the usage of \acrshort{tcp} instead of the current, unconfirmed \acrshort{udp} scheme, in \gls{nbiot}.
The proposed platform can be extended by providing the possibility to adapt the monitoring algorithm remotely. 

{
\footnotesize
\bibliographystyle{IEEEtranNMod}
\bibliography{bronnen.bib}{}
}


\vspace{-0.5cm}
{\footnotesize
\begin{IEEEbiographynophoto}{Guus Leenders}
Guus Leenders received his master’s degree summa cum laude in engineering technology at KU Leuven in 2015. He is member of the Dramco research group. There, he is involved with numerous projects in IoT. His interests are low power IoT and embedded systems.
\end{IEEEbiographynophoto}
\vspace{-0.5cm}
\begin{IEEEbiographynophoto}{Gilles Callebaut}
Gilles received the M.Sc. degree in engineering technology from KU Leuven, Belgium in 2016. In 2021, he obtained a PhD degree in engineering technology from KU Leuven. He is currently a member of the Dramco reserach group. His interests are Machine Type Communication, Internet of Things, embedded systems and everything mobile.
\end{IEEEbiographynophoto}
\vspace{-0.5cm}
\begin{IEEEbiographynophoto}{Liesbet Van der Perre}
Liesbet Van der Perre received her PhD degree in Electrical Engineering from KU Leuven, Belgium. She is  professor at Dramco, KU Leuven. Her research interest are energy efficient wireless communication and embedded systems.
\end{IEEEbiographynophoto}
\vspace{-0.5cm}
\begin{IEEEbiographynophoto}{Lieven De Strycker}
Lieven De Strycker is professor at the Faculty of Engineering Technology, Department of Electrical Engineering, KU Leuven. He joined the Engineering Technology department of the Catholic University College Ghent, where he founded the Dramco research group at KU Leuven. 
\end{IEEEbiographynophoto}}
\end{document}